\documentstyle[11pt,amssymb,epsf,cite]{article}

\def\crampest{\medmuskip = 1mu plus 1mu minus 1mu}
\def\uncramp{\medmuskip = 4mu plus 2mu minus 4mu}
\def\ben{\begin{equation}}
\def\een{\end{equation}}

\let\a=\alpha    
    
 \let\m=\mu \let\n=\nu

\let\C=\Chi 
  \let\re=\ref
  
\def\nn{\nonumber} \def\bd{\begin{document}} \def\ed{\end{document}}
\def\ds{\documentstyle} \let\fr=\frac \let\bl=\bigl \let\br=\bigr
\let\Br=\Bigr \let\Bl=\Bigl
\let\bm=\bibitem
\let\na=\nabla
\let\pa=\partial \let\ov=\overline
\newcommand{\be}{\begin{equation}}
\newcommand{\ee}{\end{equation}}
\def\ba{\begin{array}}
\def\ea{\end{array}}
\def\ft#1#2{{\textstyle{{\scriptstyle #1}\over {\scriptstyle #2}}}}
\def\fft#1#2{{#1 \over #2}}
\def\del{\partial}
\def\vp{\varphi}
\def\sst#1{{\scriptscriptstyle #1}}
\def\oneone{\rlap 1\mkern4mu{\rm l}}
\def\td{\tilde}
\def\wtd{\widetilde}
\def\ie{\rm i.e.\ }
\def\dalemb#1#2{{\vbox{\hrule height .#2pt
        \hbox{\vrule width.#2pt height#1pt \kern#1pt
                \vrule width.#2pt}
        \hrule height.#2pt}}}
\def\square{\mathord{\dalemb{6.8}{7}\hbox{\hskip1pt}}}
\newcommand{\ho}[1]{$\, ^{#1}$}
\newcommand{\hoch}[1]{$\, ^{#1}$}
\newcommand{\bea}{\begin{eqnarray}}
\newcommand{\eea}{\end{eqnarray}}
\newcommand{\ra}{\rightarrow}
\newcommand{\lra}{\longrightarrow}
\newcommand{\Lra}{\Leftrightarrow}
\newcommand{\ap}{\alpha^\prime}
\newcommand{\bp}{\tilde \beta^\prime}
\newcommand{\tr}{{\rm tr} }
\newcommand{\Tr}{{\rm Tr} }
\def\0{{\sst{(0)}}}
\def\1{{\sst{(1)}}}
\def\2{{\sst{(2)}}}
\def\3{{\sst{(3)}}}
\def\4{{\sst{(4)}}}
\def\5{{\sst{(5)}}}
\def\6{{\sst{(6)}}}
\def\7{{\sst{(7)}}}
\def\8{{\sst{(8)}}}
\def\n{{\sst{(n)}}}
\def\cA{{{\cal A}}}
\def\cF{{{\cal F}}}
\def\tV{\widetilde V}
\def\tW{\widetilde W}
\def\tH{\widetilde H}
\def\tE{\widetilde E}
\def\tF{\widetilde F}
\def\tA{\widetilde A}
\def\im{{{\rm i}}}
\def\tY{{{\wtd Y}}}
\def\ep{{\epsilon}}
\def\vep{{\varepsilon}}
\def\R{\rlap{\rm I}\mkern3mu{\rm R}}
\def\bD{{{\bar D}}}

\def\R{\rlap{\rm I}\mkern3mu{\rm R}}
\def\bD{{{\bar D}}}
\def\R{{{\Bbb R}}}
\def\C{{{\Bbb C}}}
\def\H{{{\Bbb H}}}
\def\CP{{{\Bbb C}{\Bbb P}}}
\def\RP{{{\Bbb R}{\Bbb P}}}
\def\Z{{{\Bbb Z}}}
\def\bA{{{\Bbb A}}}
\def\bB{{{\Bbb B}}}
\def\bC{{{\Bbb C}}}
\def\bD{{{\Bbb D}}}
\def\bZ{{{\Bbb Z}}}
\def\Re{{{\frak{Re}}}}
\def\Im{{{\frak{Im}}}}
\def\cosec{{\,\hbox{cosec}\,}}

\newcommand{\tamphys}{\it Center for Theoretical Physics,
Texas A\&M University, College Station, TX 77843, USA}
\newcommand{\umich}{\it Michigan Center for Theoretical Physics,
University of Michigan\\ Ann Arbor, MI 48109, USA}
\newcommand{\upenn}{\it Department of Physics and Astronomy,
University of Pennsylvania\\ Philadelphia,  PA 19104, USA}
\newcommand{\SISSA}{\it  SISSA-ISAS and INFN, Sezione di Trieste\\
Via Beirut 2-4, I-34013, Trieste, Italy}

\newcommand{\ihp}{\it Institut Henri Poincar\'e\\
  11 rue Pierre et Marie Curie, F 75231 Paris Cedex 05}

\newcommand{\damtp}{\it DAMTP, Centre for Mathematical Sciences,
 Cambridge University\\ Wilberforce Road, Cambridge CB3 OWA, UK}
\newcommand{\itp}{\it Institute for Theoretical Physics, University of
California\\ Santa Barbara, CA 93106, USA}

\newcommand{\auth}{M. Cveti\v{c}\hoch{\dagger}, G.W. Gibbons\hoch{\sharp}
and C.N. Pope\hoch{\ddagger}}

\begin{document}
\begin{flushright}
\hfill{DAMTP-2003-75\ \ \ MIFP-03-18\ \ \ UPR-1049-T}\\
\hfill{hep-th/0308026}
\end{flushright}


\begin{center}
{ \large {\Large\bf A String and M-theory Origin for the Salam-Sezgin Model}}

\vspace{20pt}
\auth

\vspace{3pt}
{\hoch{\dagger}\upenn}

\vspace{3pt}


\vspace{3pt}
{\hoch{\sharp}\damtp}

\vspace{3pt}
{\hoch{\ddagger}\tamphys}

\vspace{30pt}

\underline{ABSTRACT}
\end{center}

    An M/string-theory origin for the six-dimensional Salam-Sezgin
chiral gauged supergravity is obtained, by embedding it as a
consistent Pauli-type reduction of type I or heterotic supergravity on
the non-compact hyperboloid ${\cal H}^{2,2}$ times $S^1$.  We can also
obtain embeddings of larger, non-chiral, gauged supergravities in six
dimensions, whose consistent truncation yields the Salam-Sezgin
theory.  The lift of the Salam-Sezgin (Minkowski)$_4\times S^2$ ground
state to ten dimensions is asymptotic at large distances to the
near-horizon geometry of the NS5-brane.

{\vfill\leftline{}\vfill
\vskip 5pt
\footnoterule
{\footnotesize \hoch{\dagger} Research supported in part by DOE grant
DE-FG02-95ER40893, NATO grant 97061, NSF grant $\phantom{xxxxx}$ 
INTO-03585 and
the Fay. R. and Eugene L.  Langberg Chair. \vskip -12pt} \vskip 14pt
{\footnotesize \hoch{\star} Research supported in full by DOE grant
DE-FG02-95ER40899 \vskip -12pt} \vskip 14pt
{\footnotesize  \hoch{\ddagger} Research supported in part by DOE
grant DE-FG03-95ER40917.\vskip  -12pt}}

\pagebreak
\setcounter{page}{1}

\vfill\eject

\section{Introduction}

    It is well known that six-dimensional gauged Einstein-Maxwell
supergravity (contained within the theories in \cite{nishsez}) admits
a supersymmetric 2-sphere compactification to four-dimensional
Minkowski spacetime \cite{salsez}.  Motivated in part by recent
phenomenological interest in six-dimensional models, there has
recently been considerable activity investigating the properties of
this remarkable model of Salam and Sezgin
\cite{quevedo1,quevedo2,glps,gibpop,ggp}.  From a theoretical point of
view, two important properties are that the theory admits a consistent
Pauli reduction on $S^2$ \cite{gibpop}, and that the supersymmetric
(Minkowski)$_4 \times S^2$ ground state is the unique non-singular
solution with maximal four-dimensional spacetime symmetry
\cite{ggp}. An important and hitherto unsolved problem is whether or
not the model can be derived from M/string-theory.  Since the model
has a scalar field with a positive potential, it is natural to
suppose, because of the results in \cite{gibbons,malnun}, that any
non-singular internal space must be non-compact.

    In this paper we shall show that one can obtain the Salam-Sezgin
theory from a dimensional reduction of ten-dimensional type I supergravity,
and that indeed the internal space is non-compact.  The construction 
can be described as follows:   

\begin{itemize}

\item[1.] Perform a Pauli dimensional reduction\footnote{The
terminology of ``Pauli reduction'' was introduced in \cite{paulipap}
to describe the situation, first envisaged by Pauli, where a
dimensional reduction on a coset space yields a lower-dimensional
theory that includes all the gauge bosons associated with the isometry
group of the coset manifold. Such reductions are generically
inconsistent, and it is only in exceptional cases, such as the one
considered here, that a consistent reduction is possible.}  of
eleven-dimensional supergravity on $S^4$, to give maximal $SO(5)$
gauged supergravity in $D=7$.  The complete details of this reduction,
including the explicit demonstration of its consistency, were given in
\cite{nasvarvan1,nasvarvan2}.

\item[2.] Take a singular limit of the seven-dimensional theory, in
which the $SO(5)$ gauge group is In\"on\"u-Wigner contracted to
$SO(4)$.  In terms of the $S^4$ reduction, this corresponds to an
infinite ``stretching'' the $S^4$ along one axis, so that it limits to
$S^3\times \R$.  The resulting seven-dimensional theory can now be
viewed as a Pauli $S^3$ reduction of type IIA supergravity
\cite{cllp,clpst}.

\item[3.] Perform a (consistent) truncation of this $SO(4)$ gauged 
supergravity, in which all the fields associated with the Ramond-Ramond
sector of the type IIA supergravity are set to zero.  The resulting
theory can be interpreted as a consistent Pauli $S^3$ reduction of 
ten-dimensional type I supergravity, or of the heterotic theory.

\item[4.] Pass from the compact $SO(4)$ form for the gauge group to
the non-compact form $SO(2,2)$.  In terms of the dimensional
reduction, this corresponds \cite{hulwar} to replacing the $S^3$
specified in its ground state by $\sum_{A=1}^4 \delta_{AB}\, \mu^A\,
\mu^B=1$, where $\mu^A$ are coordinates on $\R^4$, by the hyperboloid
${\cal H}^{2,2}$ specified by $\eta_{AB}\, \mu^A\, \mu^B=1$, where
$\eta={\rm diag}\, (1,1,-1,-1)$.  Note that the metric induced on
${\cal H}^{2,2}$ is positive definite, and that consequently even in
its ground state, for which $ds_3^2 = \sum_{A=1}^4 d\mu^A\, d\mu^A$,
the metric on ${\cal H}^{2,2}$ is inhomogeneous.  Despite the
non-compact gauging, the seven-dimensional theory still has
positive-energy kinetic terms for all its fields.

\item[5.] Perform a standard Kaluza reduction of the seven-dimensional
theory on $S^1$.  This results in a six-dimensional non-chiral
$N=(1,1)$ supergravity, whose bosonic sector comprises the metric,
the gauge fields of $SO(2,2)\times U(1)$, an additional 2-form field
strength, a 3-form field strength, and 17 scalar fields.  

\item[6.] Perform a supersymmetric (consistent) truncation of this theory,
to obtain the chiral $N=(1,0)$ gauged supergravity of the Salam-Sezgin
model.  Its bosonic sector comprises the metric, a $U(1)$ gauge field,
a 3-form field strength, and a dilaton.

\end{itemize}

    Although, in order to make contact with previous work, we have
described the procedure initially in terms of seven-dimensional
supergravities with compact $SO(5)$ and $SO(4)$ gauge groups, one can
of course begin directly with the theory obtained by reducing type I
supergravity on the non-compact space ${\cal H}^{2,2}$.  Thus, in
summary, our construction yields an embedding of the Salam-Sezgin
theory as a consistent Pauli-type reduction of type I or heterotic
supergravity on the non-compact internal space ${\cal H}^{2,2}$ times
$S^1$.  A further lifting on $S^1$ allows us to embed the Salam-Sezgin
theory in eleven-dimensional supergravity.

\section{Seven-Dimensional $SO(2,2)$ Gauged Supergravity }

\subsection{Summary of the $N=4$, $SO(5)$ gauged supergravity}  

   We begin by summarising the salient features of the the maximal 
$SO(5)$ gauged supergravity in seven dimensions.  Specifically, we
shall present the Lagrangian for the bosonic sector, together with the
supersymmetry transformation rules.  For the complete details, the 
reader is referred to \cite{perpilvan,nasvarvan2}.  

   In the notation and normalisation that we shall use, the bosonic 
sector of the Lagrangian can be written as
\crampest
\bea
{\cal L}_7 &=& R\, {*\oneone} - {* P_{ij}}\wedge P^{ij} -
\ft12 \Pi_A{}^i\, \Pi_B{}^j\, \Pi_C{}^i\, \Pi_D{}^j\, 
{* F_\2^{AB}}\wedge F_\2^{CD} - \ft12 {\Pi^{-1}}_i{}^A\,
{\Pi^{-1}}_i{}^B\, {*S_{\3\, A}}\wedge S_{\3\, B}\nn\\
&& \!\!\!\!\!\!\!\! \!\!\!\!\!
+ \fft1{2g}\, \eta^{AB}\, S_{\3\, A}\wedge D S_{\3\, B}
-\fft1{8g}\, \ep_{A C_1\cdots C_4}\,\eta^{AB}\, 
S_{\3\, B}\wedge F_\2^{C_1 C_2}\wedge F_\2^{C_3 C_4} 
-\fft{1}{g}\, \Omega - V\, {*\oneone}\,,
\eea
\uncramp
where 
\bea
&& F_{\2 A}{}^B = d A_{\1 A}{}^B + g \,  A_{\1 A}{}^C\wedge
A_{\1 C}{}^B\,,\nn\\
&& D\, S_{\3\, A} \equiv dS_{\3\, A} + 
g\, A_{\1 A}{}^B \wedge S_{\3  B}\,,\nn\\
&& V = \ft12 g^2\, (2 T_{ij}\, T_{ij} - (T_{ii})^2)\,,\qquad
T_{ij} = {\Pi^{-1}}_i{}^A\, {\Pi^{-1}}_j{}^B\, \eta_{AB}\,,\nn\\
&& {\Pi^{-1}}_i{}^A\, (\delta_A{}^B\, d + g\, A_{\1 A}{}^B)\, \Pi_B{}^k\,
\delta_{kj} = P_{ij} + Q_{ij}\,; \qquad P_{ij}=P_{(ij)}\,,\quad
Q_{ij}= Q_{[ij]}\,,\label{d7defs}
\eea
and by definition we have 
\be
A_\1^{AB} \equiv \eta^{AC}\, A_{\1 C}{}^B\,,\qquad \hbox{with}\ \ 
A_\1^{AB}= -A_\1^{BA}\,.
\ee
Here, the metric $\eta_{AB}$ for the gauge group is just $\delta_{AB}$
in the compact $SO(5)_g$ gauging.  Later, we shall consider
non-compact gaugings, for which $\eta_{AB}$ will have indefinite
signature.  Note, however, that even in the case of non-compact
gaugings, the composite group $SO(5)_c$ will retain its compact form,
with its metric $\delta_{ij}$.  The quantity $\Omega$ denotes the
Chern-Simons terms for the Yang-Mills fields, with $d\Omega \sim ({\rm
tr}\, F_\2^2)^2 + {\rm tr}\, F_\2^4$ (see \cite{perpilvan} for
details). The formulation of the supergravity theory that we are using
here is the one of \cite{perpilvan,nasvarvan2}, in which the
``vielbein'' $\Pi_A{}^i$ for the scalar coset manifold
$SL(5,\R)/SO(5)_c$ is in $SL(5,\R)$, and $Q_{ij}$ is a composite
connection.  The Yang-Mills fields transform under the gauge group
$SO(5)_g$.  The $SO(5)_c$ and $SO(5)_g$ groups of this compact case
become equivalent if one imposes the symmetric gauge $\Pi=\Pi^t$.

   For some purposes, it is useful to introduce a tensor $T^{AB}$ in
place of $T_{ij}$, in order to parameterise the scalar fields:
\be
T^{AB} \equiv {\Pi^{-1}}_i{}^A\, {\Pi^{-1}}_i{}^B\,.
\ee
In terms of $T^{AB}$, the scalar potential is given by
\be
V = \ft12 g^2\, (T^{AB}\, T^{CD}\, \eta_{AC}\, \eta_{BD} -
(T^{AB}\, \eta_{AB})^2)\,.
\ee

    The fermionic sector of the $SO(5)$ gauged theory comprises the
gravitini $\psi_\mu^I$ and gaugini $\lambda_i^I$, where the $I$ index
denotes the four-dimensional spinor of $SO(5)_c$.  Their supersymmetry
transformation rules are \cite{perpilvan}
\bea
\delta\psi_\mu &=& D_\mu\, \ep - \ft1{20}\, g\, T_{ii}\, \Gamma_\mu\, \ep 
- \ft1{40\sqrt2}\, (\Gamma_\mu{}^{\nu\rho} - 8 \delta_\mu^{\nu}\, 
\Gamma^\rho)\, \gamma_{ij}\, \ep\, \Pi_A{}^i\, \Pi_B{}^j\, F_{\nu\rho}^{AB}
\nn\\
&& - \ft{1}{60}\, (\Gamma_\mu{}^{\nu\rho\sigma} - \ft92 \delta_\mu^\nu\,
\Gamma^{\rho\sigma})\, \gamma^i\, \ep\, {\Pi^{-1}}_i{}^A\, 
S_{\nu\rho\sigma\,, A}\,,\nn\\
\delta \lambda_i &=& \ft1{16\sqrt2}\, \Gamma^{\mu\nu}\, 
(\gamma_{k\ell}\, \gamma_i - \ft15 \gamma_i\, \gamma_{k\ell})\, \ep\, 
\Pi_A{}^k\, \Pi_B{}^\ell\, F_{\mu\nu}^{AB} - 
\ft{1}{120}\, \Gamma^{\mu\nu\rho}\, (\gamma_i{}^j - 4 \delta_i^j)\, 
\ep\, {\Pi^{-1}}_j{}^A\, S_{\mu\nu\rho\,, A} \nn\\
&& + \ft12g\, (T_{ij} - \ft15 T_{kk}\, \delta_{ij})\, \gamma^j\, \ep +
\ft12 \Gamma^\mu\, \gamma^j\, \ep\, P_{\mu\,  ij}\,.\label{susyf}
\eea
Here $\gamma_i$ denotes the ``internal'' Dirac matrices of $SO(5)_c$, and
the associated spinor indices $I,J,\ldots$ have been suppressed.  The spinors
$\lambda_i$ are subject to the constraint
\be
\gamma^i\, \lambda_i=0\,.\label{con1}
\ee
The 
seven-dimensional Dirac matrices are denoted by $\Gamma_\mu$.  The covariant
derivative $D_\mu$ acting on spinors is given by
\be
D\, \ep = d\, \ep + \ft14 \omega_{ab}\, \Gamma^{ab}\, \ep + 
\ft14 Q_{ij}\, \gamma^{ij}\, \ep\,.
\ee
The supersymmetry transformation rules for the bosonic fields are given by
\bea
\delta e^a_\mu &=& \ft12 \bar\ep\, \Gamma^a\, \psi_\mu\,,\nn\\
\Pi_A{}^i\, \Pi_B{}^j\, \delta A_\mu^{AB} &=& \ft1{2\sqrt2}\, 
\bar\ep\, \gamma^{ij}\, \psi_\mu  + \ft1{4\sqrt2}\, \bar\ep\, \Gamma_\mu\,
\gamma^k\, \gamma^{ij}\, \lambda_k\,,\nn\\
{\Pi^{-1}}_i{}^A\, \delta \Pi_A{}^j &=& \ft14 (\bar\ep\, \gamma_i\, 
\lambda^j + \bar\ep\, \gamma^j\, \lambda_i)\,,\label{susyb}\\
\delta S_{\mu\nu\rho, A} &=& -\ft{3}{4\sqrt2}\, \Pi_A{}^i\, 
(2\bar\ep\, \gamma_{ijk}\, \psi_{[\mu} + \bar\ep\, \Gamma_{[\mu} \,
\gamma^\ell\, \gamma_{ijk}\, \lambda_\ell)\, \Pi_B{}^j\, \Pi_C{}^k\, 
F_{\nu\rho]}^{BC} \nn\\
&& - \ft{3}{2}\, \delta_{ij}\, \Pi_A{}^j\, D_{[\mu}\, (2\bar\ep\, 
\Gamma_{\nu}\, \gamma^i\, \psi_{\rho]} + \bar\ep\, \Gamma_{\nu\rho]}\, 
\lambda^i) \nn\\
&&+ \ft{1}{2}\, \delta_{AB}\, {\Pi^{-1}}_i{}^B\, 
(3 \bar\ep\, \Gamma_{[\mu\nu}\, \gamma^i\, \psi_{\rho]} -
\bar\ep\, \Gamma_{\mu\nu\rho}\, \lambda^i)\,.\nn
\eea
 
\subsection{The $N=2$ gauged $SO(4)$ limit of $N=4$, $D=7$ supergravity}

   We next consider the In\"on\"u-Wigner group contraction limit of
the $SO(5)$ gauged theory in seven dimensions, to obtain a maximal
supergravity with $SO(4)$ gauging.  The procedure was described in a
truncated system in \cite{cllp}, and implemented in the bosonic sector
of the full theory in \cite{clpst}.  The idea is to decompose the
$SO(5)$ vector indices in a $4+1$ split, and make appropriate
rescalings of the resulting $SO(4)$-valued fields, such that in a
singular limit the gauge group degenerates to $SO(4)$.  For the
$SO(5)_c$ and $SO(5)_g$ indices we make the decompositions
\be
i = (0, \a)\,,\qquad A=(0,\bar A)\label{decomps}
\ee
respectively, where $\a$ and $\bar A$ run over the values $1,2,3,4$.

    Following \cite{clpst}, we decompose and rescale the bosonic fields,
and the gauge coupling constant, as follows:
\bea
&&g = k^2\, \td g\,,\qquad A_\1^{0\bar A} = k^3\, \wtd A_\1^{0\a}
\,\qquad A_\1^{\bar A\bar B} = k^{-2}\, \wtd A_\1^{\bar A\bar B}\,,\nn\\
&& S_\3^0 = k^{-4}\, \wtd H_\3\,,\qquad 
S_\3^{\bar A} = k\, \wtd S_\3^{\bar A}\,,\nn\\
&&T_{ij} = \pmatrix{ k^8\, \Phi^{-1} + k^8\, \chi_\gamma\, \chi^\gamma
 & - k^3\, \chi^\a \cr
- k^3\, \chi^\a & k^{-2}\,\Phi^{1/4}\,  M_{\a\beta}}\,,
\eea
where $M_{\a\beta}$, like the original scalar matrix $T_{ij}$, is
unimodular.  In the fermionic sector, no scalings by $k$ are required.

    After taking the limit $k\longrightarrow 0$, one obtains the 
group-contracted $SO(4)$-gauged theory.  The complete results for
the bosonic sector can be found in \cite{clpst}.  Note that after
talking the limit, $\wtd H_\3$, which is the rescaled 3-form $S_\3^0$, 
acquires the interpretation of a field-strength for a 2-form potential,
rather than being a fundamental field in its own right (see \cite{clpst}).

    At this point no truncation of fields has been performed, and the
$SO(4)$-gauged theory can be interpreted as a consistent Pauli
reduction of type IIA supergravity on $S^3$ \cite{clpst}.  It has
$N=4$ (\ie maximal) supersymmetry.  Since the seven-dimensional theory
at this stage is quite complicated, and we shall not require its
detailed form for our purposes, we shall not repeat expressions found
in \cite{clpst}, and their fermionic counterparts, here.  Rather, our
interest at this stage is in performing a consistent truncation of the
seven-dimensional theory to one with $N=2$ supersymmetry, which can be
interpreted as a consistent Pauli reduction of ten-dimensional type I
supergravity, or equivalently, the heterotic theory.

    To make this truncation, in the bosonic sector we set the four
scalars $\chi^\a$, the four gauge bosons $A_\1^{0\a}$ and the four
3-form fields $S_\3^\a$ to zero.  It is easily seen from the formulae
in \cite{clpst} that this truncation is a consistent one, in the sense
that it is compatible with the equations of motion for the truncated
fields.  Of course since we wish to obtain a supersymmetric truncated
theory it is also necessary to set the appropriate fermionic
superpartners to zero, and to check that this fermionic truncation is
consistent with the supersymmetry transformation rules.

    We find that the appropriate fermionic truncation can be achieved by
first projecting all the spinors into eigenstates of $\gamma_0$, which
is the chirality operator with respect to the $SO(4)$ contracted subgroup 
of the original $SO(5)$ internal group.  Writing $\ep = \ep^+ + \ep^-$,
$\ep^\pm = \pm \gamma_0\, \ep^\pm$, etc, we then make the following
truncation:
\be
\ep^-=0\,,\qquad \psi_\mu^- = 0\,,\qquad 
\lambda_0^- =0\,, \qquad \lambda_\a^+ =0\,.
\ee
The constraint (\ref{con1}) now leads to the relation
\be
\lambda_0^+ = - \gamma^\a\, \lambda_\a^-\,,\label{rel1}
\ee
and so the independent fermionic fields that survive the truncation 
can be taken to be just $\psi_\mu^+$ and $\lambda_\a^-$.

   After some algebra, it can be verified from the original 
transformation rules (\ref{susyf}) and (\ref{susyb}) that
the bosonic and fermionic truncations described above are fully
consistent with supersymmetry.

    At this stage, it is useful to summarise the details of the 
bosonic contraction and truncation, in terms of the scalar vielbein
$\Pi_A{}^i$.  Thus we have
\bea
\Pi_0{}^0 &=& k^{-4}\, \Phi^{1/2}\,,\qquad \Pi_{\bar A}{}^\a = k\, 
\Phi^{-1/8}\, \pi_{\bar A}{}^\a\,,\nn\\
g &=& k^2\, \td g\,,\qquad A_\1^{\bar A\bar B}=k^{-2}\, 
\wtd A_\1^{\bar A \bar B}\,,
\qquad S_\3 = k^{-4}\, \wtd H_\3\,,\nn\\
A_\1^{0\bar A} &=& 0\,,\qquad
S_\3^{\bar A} = 0\,,
\eea
where $\det(\pi_{\bar A}{}^\a)=1$.  As described in \cite{clpst},
although $S_\3^0$ was itself a fundamental field, subject to an
``odd-dimensional self-duality equation'' in the original
$SO(5)$-gauged theory, it now acquires the interpretation of being a
3-form field strength for a 2-form potential in the contraction limit.
In what follows we shall drop the tildes that were previously used to
distinguish between the original fields and the rescaled fields of the
In\"on\"u-Wigner contraction limit.  The Lagrangian for the bosonic
sector of the $N=2$ gauged $SO(4)$ supergravity is thus given by (see
\cite{clpst})
\bea
{\cal L}_7 &=& R\, {*\oneone} - \ft5{16}\, \Phi^{-2}\, {*d\Phi}\wedge d\Phi 
- {* p_{\a\beta}}\wedge p^{\a\beta} - \ft12 \Phi^{-1}\, 
{* H_\3}\wedge  H_\3 \nn\\
&& - \ft12 \Phi^{-1/2}\, 
\pi_{\bar A}{}^\a \, \pi_{\bar B}{}^{\beta}\, \pi_{\bar C}{}^\a\, 
\pi_{\bar D}{}^\beta\, 
{* F_\2^{\bar A\bar B}}\wedge F_\2^{\bar C \bar D}
 - \fft1{g} \,  \Omega -  V \, {*\oneone}\,,\label{d7n2lag}
\eea
where $p_{\a\beta}$ is defined analogously to $P_{ij}$ in
(\ref{d7defs}), namely 
\be
p_{\a\beta} = {\pi^{-1}}_{(\a}{}^{\bar A}\,
[\delta_{\bar A}{}^{\bar B}\, d + g\, A_{\1\bar A}{}^{\bar B}]\, 
\pi_{\bar B}{}^\gamma\, \delta_{\beta)\gamma}\,.
\ee
Thus $p_{\a\beta}$ 
is traceless, and it describes the derivatives of the 9 scalar fields in the 
unimodular $M_{\a\beta}$.    The scalar potential is given by
\be
V = \ft12 g^2\, \Phi^{1/2}\, ( 2 M_{\a\beta}\, M_{\a\beta} 
-(M_{\a\a})^2)\,.\label{mpot}
\ee
Note that the invariant tensor $\eta_{\bar A\bar B}$ for the group of the
gauge symmetry appears in the expressions
\be
M_{\a\beta} = {\pi^{-1}}_\a{}^{\bar A}\, 
{\pi^{-1}}_\beta{}^{\bar B}\,  \eta_{\bar A \bar B}\,,\qquad
A_{\1}^{\bar A \bar B} = \eta^{\bar A\bar C}\, 
A_{\1 \bar C}{}^{\bar B}\,.
\ee
For now, since we are discussing the compact case with $SO(4)$ gauging,
we have $\eta_{\bar A\bar B}=\delta_{\bar A\bar B}$.  Note that the 
$\a,\beta$ indices are always raised and lowered with a Kronecker delta,
regardless of whether the gauge group is compact or non-compact.

   We find that the supersymmetry transformation rules for the fermions in 
the $N=2$ gauged $SO(4)$ theory are given by
\bea
\delta\psi_\mu &=& D_\mu\, \ep - \ft1{20}\, g\, M_{\a\a}\, 
 \Phi^{1/4}\, \Gamma_\mu\, \ep - \ft1{40\sqrt2}\, 
(\Gamma_\mu{}^{\nu\rho} - 8\delta_\mu^\nu\, \Gamma^\rho)\, \gamma_{\a\beta}\,
\ep\, \Phi^{-1/4}\, 
\pi_{\bar A}{}^\a\, \pi_{\bar B}{}^\beta\, F_{nu\rho}^{\bar A\bar B}
\nn\\
&& - \ft{1}{60}\, (\Gamma_\mu{}^{\nu\rho\sigma} - \ft92 
\delta_\mu^\nu\, \Gamma^{\rho\sigma})\, \ep\, \Phi^{-1/2}\, 
H_{\nu\rho\sigma}\,,\nn\\
\delta \lambda_\a &=& \ft12 \Gamma^\mu\, \gamma^\beta\, \ep\, 
P_{\mu, \a\beta}
+ \ft1{16\sqrt2}\, \Gamma^{\mu\nu}\, (\gamma_{\beta\gamma}\, \gamma_\a
- \ft15 \gamma_\a\, \gamma_{\beta\gamma})\, \ep\, \Phi^{-1/4}\, 
\pi_{\bar A}{}^\beta\, \pi_{\bar B}{}^\gamma\, 
F_{\mu\nu}^{\bar A \bar B} + \nn\\
&&- \ft{1}{120}\, \Gamma^{\mu\nu\rho}\, \gamma_\a\, \ep\, \Phi^{-1/2}\, 
H_{\mu\nu\rho} + \ft12 g\, (M_{\a\beta} -\ft15  M_{\gamma \gamma}
\, \delta_{\a\beta})\, \Phi^{1/4}\, \gamma^\beta\, \ep\,,\label{susyf2}
\eea
where we are now suppressing the internal $SO(4)$ chirality superscripts on
$\ep^+$, $\psi_\mu^+$ and $\lambda_\a^-$.   Here $P_{\a\beta} = 
p_{\a\beta} - \ft18 \Phi^{-1}d\Phi\, \delta_{\a\beta}$.

    The supersymmetry transformations rules for the bosonic fields
take the form
\bea
&&\delta e^a_\mu =  \ft12 \bar\ep\, \Gamma^a\, \psi_\mu\,,\nn\\
&&\pi_{\bar A}{}^\a\, 
\pi_{\bar B}{}^\beta \, \delta A_\mu^{\bar A\bar B} 
= \ft1{2\sqrt2}\, 
\bar\ep\, \gamma^{\a\beta}\, \psi_\mu  + 
\ft1{2\sqrt2}\, \bar\ep\, \Gamma_\mu\,(
\gamma^\beta\,  \lambda^\a - \gamma^\a\, \lambda^\beta)\,,\nn\\
&&{\pi^{-1}}_\a{}^{\bar A}\, 
\delta \pi_{\bar A}{}^\beta = \ft14 (\bar\ep\, \gamma_\a\, 
\lambda^\beta + \bar\ep\, \gamma^\beta\, \lambda_\a)\,,
\qquad \Phi^{-1}\, \delta \Phi = - \bar\ep\, \gamma^\a\, \lambda_\a\,,
\label{susyb2}\\
&&\delta H_{\mu\nu\rho} = -\ft{3}{2\sqrt2}\, \Phi^{1/4}\, 
(\bar\ep\, \gamma_{\a\beta}\, \psi_{[\mu} - \bar\ep\, \Gamma_{[\mu} \,
\gamma_{\a\beta\gamma}\, \lambda^\gamma)\, \pi_{\bar B}{}^\a\, 
\pi_{\bar C}{}^\beta\, 
F_{\nu\rho]}^{\bar B\bar C} \nn\\
&&\qquad\qquad   - \ft{3}{2}\, \Phi^{1/2}\, 
D_{[\mu}\, (2\bar\ep\, 
\Gamma_{\nu}\, \psi_{\rho]} + \bar\ep\, \Gamma_{\nu\rho]}\, 
\gamma^\a\, \lambda_\a) 
+ \ft{1}{2}\, \Phi^{-1/2}\, 
(3 \bar\ep\, \Gamma_{[\mu\nu}\, \psi_{\rho]} +
\bar\ep\, \Gamma_{\mu\nu\rho}\,\gamma^\a\,\lambda_\a)\,.\nn
\eea

\subsection{The $N=2$ gauged $SO(2,2)$ theory in $D=7$}\label{so22sec}

   Our discussion so far in this section has focussed on the $N=4$
compact $SO(5)$ gauging in seven dimensions, its contraction limit to
an $SO(4)$ gauged supergravity, again with $N=4$, and then the
truncation to an $N=2$ gauged theory, again with $SO(4)$ Yang-Mills
fields.  In this subsection, we turn to the consideration of a
non-compact gauging for the $N=2$ theory, with $SO(2,2)$ as gauge
group. (See \cite{hulwar}, and references therein, for a discussion of
the higher-dimensional origins of non-compact gaugings in
supergravities.)

    The main idea is to take $\eta_{\bar A\bar B} = {\rm diag}\,
(+1,+1,-1,-1)$.  The gauge group becomes $SO(2,2)$, but the global
$SO(4)$ group is reduced to its intersection with the gauge group,
namely $SO(2)\times SO(2)$.  The internal manifold is the hyperboloid
${\cal H}^{2,2}$, specified by 
\be
\mu_1^2 + \mu_2^2 -\mu_3^2 -\mu_4^2 =1\,,\label{surfacex}
\ee
embedded in Euclidean space ${\Bbb E}^4$ with the standard metric 
$ds^2 = d\mu_1^2 + d\mu_2^2 + d\mu_3^2 + d\mu_4^2$.  The hyperboloid is
invariant under the action of the non-compact group $SO(2,2)$, and may
be identified with the symmetric space $SO(2,2)/SO(2,1)$.  However, 
the metric induced from the Euclidean metric is not the standard 
homogeneous metric on $SO(2,2)/SO(2,1)$, which has signature 
$(1,2)$, but rather an inhomogeneous metric of cohomogeneity one,
whose isometry group is $SO(2)\times SO(2)$.  Concretely, it
is convenient to parameterise the ${\Bbb E}^4$ coordinates as
\be
\mu^1 + \im\, \mu^2 = \cosh\rho\, e^{\im\,\a}\,,\qquad
\mu^3 + \im\, \mu^4 = \sinh\rho\, e^{\im\, \beta}\,,\label{coords}
\ee
where $0\le \rho <\infty$, $0\le \a < 2\pi$, $0\le \beta < 2\pi$, 
so that the constraint (\ref{surfacex}) is satisfied.  

   In the ground state, where $M_{\bar A\bar B}=\delta_{\bar A \bar
B}$, the metric on ${\Bbb E}^4$ induces the metric
\be
ds_3^2 = \cosh 2\rho\, d\rho^2 + \cosh^2\rho\, d\a^2 + 
\sinh^2\rho\, d\beta^2\label{h22met}
\ee
on ${\cal H}^{2,2}$.  Because the length of the $\a$ circle never
vanishes, the topology of ${\cal H}^{2,2}$ is $\R^2\times S^1$, where
$\rho$ and $\beta$ parameterise the $\R^2$ and $\a$ parameterises
the $S^1$.

    By inspection, the Lagrangian (\ref{d7n2lag}), with this choice
$\eta_{\bar A\bar B} = {\rm diag}\, (1,1,-1,-1)$, has $SO(2,2)$ 
local gauge invariance and positive kinetic energies for all the 
fields.  For the scalars, this is because the $\a$ and $\beta$ 
indices in $-{*P_{\a\beta}}\wedge P^{\a\beta}$ are always raised and lowered
with the positive-definite metric $\delta_{\a\beta}$, whether or not
the gauge group is compact.  Likewise, the gauge-field kinetic energies
are all positive, as can be seen by defining $F_\2^{\a\beta} \equiv
\pi_{\bar A}{}^\a\, \pi_{\bar B}{}^\beta\, F_\2^{\bar A\bar B}$, so
that one has $-\ft12 \Phi^{-1/2}\, {*F_\2^{\a\beta}}\wedge F_\2^{\a\beta}$. 

   In a gauge theory without scalars, it is not possible to construct
gauge-invariant kinetic terms for the gauge bosons that have positive
kinetic energy.  It is the presence of the scalar fields $\pi_{\bar
A}^i$, transforming non-trivially under the gauge group, that allows
the existence of gauge-invariant kinetic terms of positive energy.
Under this gauge transformation, the scalar kinetic term is unchanged
because $P_{\a\beta}$ is invariant.  Of course in the ground state,
where $M_{\bar A\bar B}= \delta_{\bar A\bar B}$, the non-compact gauge
group $SO(2,2)$ is broken down to its maximal compact subgroup,
$SO(2)\times SO(2)$.

\section{Kaluza Reduction to Six Dimensions}\label{kaluzasec} 

   In this section, we shall show how the $N=(1,0)$ chiral
Einstein-Maxwell gauged supergravity in six dimensions can be obtained
by performing a Kaluza reduction of the seven-dimensional $N=2$ gauged
$SO(2,2)$ theory of section \ref{so22sec} to give an $N=(1,1)$
supergravity, and then performing a consistent chiral truncation of
this non-chiral theory.  We begin by summarising the general
formalism for the Kaluza reduction, and then we implement it, together
with the chiral truncation, in the subsequent subsection.

\subsection{Kaluza reduction to the $N=(1,1)$ theory}\label{d611sec}

   The procedure for performing a Kaluza reduction on $S^1$ is well
established, and here we shall just review the essential points, in
order to establish our notation.  We shall now place hats on the
seven-dimensional fields, and take the seven-dimensional coordinate and
tangent-frame indices to be $\hat\mu=(\mu,z)$ and $\hat a =(a,7)$ 
respectively. 

   The metric is reduced according to
\be
d\hat s_7^2 = e^{2\a\varphi}\, ds_6^2 + e^{-8\a\varphi}\, 
(dz + {\cal A}_\1)^2\,,
\ee
where $\a= 1/(2\sqrt{10})$, for which we choose the natural vielbein basis
\be
\hat e^a = e^{\a\varphi}\, e^a\,,\qquad \hat e^7 = e^{-4\a\varphi}\, 
(dz + {\cal A}_\1)\,.
\ee
The dilaton couplings in the above are chosen so that an
Einstein-Hilbert Lagrangian in $D=7$ reduces to give an
Einstein-Hilbert term in $D=6$, and so that the kinetic term for
$\varphi$ has its canonical normalisation.  A $p$-form potential is
reduced according to $\hat A_{\sst(p)} = A_{\sst (p)} + A_{\sst
(p-1)}\wedge dz$.  The associated field strength is reduced according
to
\be
\hat F_{\sst (p+1)} = F_{\sst (p+1)} + F_{\sst (p)}\wedge (dz +
{\cal A}_\1)\,,
\ee
where the lower-dimensional field strengths are defined by
\be
F_{\sst (p+1)} = d A_{\sst (p)} - dA_{\sst (p-1)}\wedge {\cal A}_\1\,,
\qquad F_{\sst (p)} = dA_{\sst (p-1)}\,.
\ee

   The Kaluza reduction of the fermions is determined by the requirement
that the lower-dimensional kinetic terms, like those in the higher
dimension, should have no scalar prefactors.  Thus we take
\be
\hat \lambda = e^{-\ft12 \a\varphi}\, \lambda\,,\qquad
\hat \psi_a =  e^{-\ft12 \a\varphi}\, \psi_a\,,\qquad
\hat \psi_7 =  e^{-\ft12 \a\varphi}\, \psi_7\,.
\ee
(Note that the reduction for the gravitino is expressed in terms of
vielbein components for the vector index.)  In order to get a canonical
transformation rule for the lower-dimensional gravitino, $\delta\psi_a =
\nabla_a\, \ep +\cdots$, we should reduce $\hat\ep$ according to
\be
\hat\ep = e^{\ft12\a\varphi}\, \ep\,.
\ee

    Applying the above reduction procedures to the $N=2$ gauged $SO(2,2)$
theory obtained in section \ref{so22sec} is a purely mechanical 
exercise, and since the complete details of the resulting non-chiral
$N=(1,1)$ supergravity in six dimensions are not needed for our 
present purposes, we shall just give an outline of the result here.
We find the bosonic Lagrangian in six dimensions is given by
\bea
{\cal L}_6 &=& R\, {*\oneone} - \ft5{16}\, \Phi^{-2}\, {*d\Phi}\wedge d\Phi 
- {* P_{\a\beta}}\wedge P^{\a\beta} -\ft12 {*d\varphi}\wedge
d\varphi -\ft12 e^{-10\a\varphi}\, {*{\cal F}_\2}\wedge {\cal F}_\2 \nn\\
&&- \ft12 \Phi^{-1}\, e^{-4\a\varphi}\,  
{*H_\3}\wedge H_\3  - \ft12 \Phi^{-1}\, e^{6\a\varphi}\,  
{*H_\2}\wedge H_\2 -\fft{1}{g}\, \Omega \label{n11bos6}\\
&&- \ft12 \Phi^{-1/2}\,  
\pi_{\bar A}{}^\a \, \pi_{\bar B}{}^{\beta}\, \pi_{\bar C}{}^\a\, 
\pi_{\bar D}{}^\beta\, \Big( e^{-2\a\varphi}\, 
{* F_\2^{\bar A\bar B}}\wedge F_\2^{\bar C \bar D} + e^{8\a\varphi}\, 
{* F_\1^{\bar A\bar B}}\wedge F_\1^{\bar C \bar D}\Big)
 -  V \, {*\oneone}\,,\nn
\eea
where the scalar potential is now given by
\be
V = \ft12 g^2\, \Phi^{1/2}\, e^{2\a\varphi}\, ( 2 M_{\a\beta}\, M_{\a\beta} 
-(M_{\a\a})^2)\,.\label{mpot6}
\ee
The fields $H_\2$ and $F_\1^{\bar A\bar B}$ come from the reductions
of $H_\3$ and $F_\2^{\bar A\bar B}$ respectively.

    The associated six-dimensional fermionic fields are $\psi_\m$, 
$\psi_7$, and $\lambda_\a$, still carrying suppressed internal $SO(4)$
spinor indices as well as Spin$(1,5)$ spacetime spinor indices.  The 
internal $SO(4)$ chiralities, as in $D=7$, are $\gamma_0\, \psi_\mu =
+\psi_\mu$, $\gamma_0\, \psi_7 = + \psi_7$, and $\gamma_0\, \lambda_\a =
-\lambda_\a$.  Of course, as usual in a dimensional reduction, it is
convenient to make a redefinition of the gravitino, of the form
${\psi_\mu}' = \psi_\mu - \ft14 \Gamma_\mu\, \psi_7$ here, in order to 
obtain diagonalised kinetic terms for the gravitini ${\psi_\mu}'$ and the 
spin-$\ft12$ fields $\psi_7$ and $\lambda_\a$.  

    The supersymmetry transformation rules for the six-dimensional
fields can be straightforwardly read off by applying the Kaluza
reduction procedure described above to the transformation rules in
section \ref{so22sec}.

\subsection{The truncation to the Salam-Sezgin $N=(1,0)$ theory}

    We are now in a position to implement the final stage of our
reduction procedure, in which we perform a truncation of the six-dimensional
$N=(1,1)$ supergravity described in section \ref{d611sec} to a chiral
$N=(1,0)$ supergravity.  Of course a crucial point about this truncation,
as with the previous ones we have implemented, is that it must be 
consistent with both the equations of motion and the supersymmetry 
transformation rules of the fields that are being set to zero.

   In the bosonic sector, the truncation consists of setting to zero
the 9 scalar fields parameterised by $\pi_{\bar A}{}^\a$; the
Kaluza-Klein vector ${\cal A}_\1$; the 2-from $H_\2$; the 1-form field
strengths $F_\1^{\bar A\bar B}$; one combination of the
seven-dimensional scalar field $\Phi$ and the Kaluza-Klein scalar
$\varphi$; and, finally, setting to zero all except one vector within
the $SO(2,2)$ Yang-Mills sector. Thus we set
\bea
&&\pi_{\bar A}{}^\a = \delta_{\bar A}{}^\a\,,\qquad
{\cal A}_\1=0\,,\qquad F_\1^{\bar A\bar B}=0\,,\qquad
\Phi=e^{16\a\varphi} = e^{-\ft45\phi}\,,\nn\\
&&A_\1^{12} = - A_\1^{34} = \ft1{2}\, A_\1\,.
\eea
We also, for convenience, define a rescaled gauge coupling, 
\be
\bar g = \fft{g}{\sqrt2} \,.\label{grescale}
\ee

     We first note that this truncation leads to the six-dimensional 
bosonic Lagrangian
\be
{\cal L}_6 = R\,{*\oneone} - \ft14{*d\phi}\wedge d\phi - 
\ft12 e^{\ft12\phi}\, {*F_\2}\wedge F_\2 -\ft12 e^\phi\, {*H_\3}\wedge H_\3
- 8 \bar g^2\, e^{-\ft12\phi}\, {*\oneone}\,,\label{d6boslag}
\ee
where $dH_\3 = \ft12 F_\2\wedge F_\2$.  It is straightforward to verify
that this truncation is indeed consistent with the bosonic equations of
motion.  The Lagrangian (\ref{d6boslag}) precisely describes the 
bosonic sector of the six-dimensional Salam-Sezgin gauged Einstein-Maxwell
supergravity.\footnote{The bosonic sector has also been obtained via
a generalised dimensional reduction of ungauged seven-dimensional 
supergravity \cite{kelu}.  It is not clear whether that construction
would allow an extension to include the fermionic sector.}

    At the same time as truncating the bosonic sector, we must also
set to zero appropriate fermionic fields, in order to obtain a
supersymmetric $N=(1,0)$ theory.  To do this, we first decompose the
six-dimensional fermionic fields into eigenstates of $\Gamma_7$,
which plays the role of the chirality operator in $D=6$.  Thus we
write $\ep= \ep^+ + \ep^-$, where $\Gamma_7\, \ep^\pm = \pm \ep^\pm$,
etc.  Note that these chiralities are quite independent of the $SO(4)$
internal chiralities under $\gamma_0$, which we introduced in the
truncation to $N=2$ supersymmetry in $D=7$.  We then suppressed the
$SO(4)$ chirality labels, in order to avoid a proliferation of $\pm$
superscripts in this final stage of the construction.

   We find that the appropriate truncation in the fermionic sector
is obtained by setting
\bea
&&\ep^-=0\,,\qquad
\psi_\mu^-=0\,,\qquad \psi_7^+=0\,,\qquad \lambda_\a^- = \gamma_\a\, 
\lambda^-\,,\qquad \lambda_\a^+ = \eta_{\a\beta}\, \gamma_\beta\, 
\lambda^+\,,\nn\\
&&\psi_7^- = -2 \lambda^- \,.
\eea
Thus the remaining independent fermionic degrees of freedom are
described by the fields $(\psi_\mu^+, \lambda^-, \lambda^+)$, and the
supersymmetry parameter is $\ep^+$.  As well as their
explicitly-indicated six-dimensional chiralities, they are also
subject to the internal chirality constraints
\be
\gamma_0\, \psi^+_\mu = +\psi^+_\mu\,,\qquad 
\gamma_0\, \chi^- = -\chi^-\,,\qquad \gamma_0\, \lambda^+ = -\lambda^+\,.
\ee

As before, to obtain diagonal kinetic terms we should define
\be
{\psi_\mu^+} = {\psi_\mu^+}' + \ft14 \Gamma_\mu\, \psi_7^-\,.
\ee
It is convenient also to introduce rescaled fermionic fields, which will
have canonically-normalised kinetic terms.  Tracing back through the
sequence of reductions and truncations, we find that the original
fermion kinetic terms ${\cal L}_F = \bar\lambda^i \, \Gamma^\mu\,
D_\mu\, \lambda_i + \bar\psi_\mu \, \Gamma^{\mu\nu\rho}\ D_\nu\,
\psi_\rho$ of the seven-dimensional $SO(5)$ gauged theory
\cite{perpilvan} give rise, after our reduction and truncation to the
$N=(1,0)$ Salam-Sezgin supergravity, to the six-dimensional fermionic
kinetic terms ${\cal L}_F = 25\bar \lambda^-\, \Gamma^\mu\, D_\mu\,
\lambda^- + 4 \bar\lambda^+\, \Gamma^\mu\, D_\mu\, \lambda^+ +
{\bar\psi_\mu^+}{}'\, \Gamma^{\mu\nu\rho}\, {\psi_\rho^+}'$.  Thus if we
define
\be
\lambda^- =\ft15 \chi\,,\qquad \lambda^+ =\ft12 \lambda\,,
\ee
then $({\psi_\mu^+}', \chi, \lambda)$ describe the canonically-normalised
fermionic fields of the Salam-Sezgin theory.

    It is now a straightforward matter to show that the truncation of
the bosons and fermions described above is consistent with the
supersymmetry transformation rules that descend from seven dimensions.
It is worth remarking that two crucial ingredients in establishing
the consistency are that
\be
M_{\a\a}=0\,,\qquad  F_\2^{\a\beta}\, \gamma_{\a\beta}\, \ep^+=0\,.
\ee
The first of these equations follows because we made the transition
to the $SO(2,2)$ non-compact gauging, whilst the second follows from
the internal $SO(4)$ chirality condition $\gamma_0\, \ep^+ = \ep^+$,
which implies that $\gamma_{12}\, \ep^+ = \gamma_{34}\, \ep^+$ (
we choose conventions where $\gamma_0=-\gamma_{1234}$).

   We find that the supersymmetry transformation rules for the fermions
that remain in the truncated theory are given by 
\bea
\delta {\psi^+_\mu}' &=& D_\mu\, \ep^+  + \ft1{48} e^{\ft12\phi}\, 
H_{\nu\rho\sigma}\, \Gamma^{\nu\rho\sigma}\, \Gamma_\mu\, \ep^+ 
  \,,\nn\\
\delta \chi &=&  \ft1{4} [ \del_\mu\phi\, \Gamma^\mu 
   -\ft16 e^{\ft12\phi}\, \Gamma^{\mu\nu\rho}\, H_{\mu\nu\rho}]\,
\ep^+ \,,\label{ssferm}\\
\delta \lambda &=& -\ft1{4\sqrt2} [e^{\ft14\phi}\, F_{\mu\nu}\, 
\Gamma^{\mu\nu}\, \gamma_{12} - 8 \bar g\, e^{-\ft14\phi}]\, \ep^+
\,,\nn
\eea
where
\be
D_\mu\ep^+ = \nabla_\mu\ep^+ + \bar g\, A_{\mu}\, \gamma_{12}\, \ep^+\,.
\label{d6cov}
\ee
The transformation rules (\ref{ssferm}) are precisely those for the
fermions in the Salam-Sezgin theory.  We see from (\ref{d6cov}) that
the fermions all carry charge $\bar g$ under the $U(1)$ gauge field
$A_\mu$.  One can pass to a complex notation, in which one takes
$\gamma_{12}\, \ep^+ = -\im\, \ep^+$, or else keep $\gamma_{12}$
explicitly, which corresponds to working with a 2-component chiral
$SO(4)$ representation for symplectic Majorana spinors in $D=6$.

    Turning now to the supersymmetry transformations of the bosonic
fields, we find
\bea
\delta e^a_\mu &=& \ft12 \bar\ep^+\, \Gamma^a\, {\psi^+_\mu}'
                       - \ft1{20} \bar\ep^+\, \Gamma^a{}_\mu\, \chi\,,\nn\\
\delta\phi &=& \bar\ep^+\, \chi\,,\nn\\
\delta A_\mu &=& -\ft1{\sqrt2}\, e^{-\ft14\phi}\, \bar\ep^+\, \Gamma_\mu\, 
\gamma_{12}\, \lambda\,.
\eea
The second term on the right-hand side of the vielbein transformation
rule can be removed by performing a compensating local Lorentz
transformation, with parameters $\Lambda^a{}_b = \ft1{20}\,
\bar\ep^+\, \Gamma^a{}_b\, \chi$.  It is also straightforward to write
down the transformation rule for the 3-form $H_{\mu\nu\rho}$, and from this,
one can deduce the transformation rule for $B_{\mu\nu}$.  

\newpage
\section{Embedding of the Salam-Sezgin Theory in Ten Dimensions}

    Now that we have established how the Salam-Sezgin theory can be 
embedded into a non-compact gauged supergravity in seven dimensions, 
we can straightforwardly lift it back to ten dimensions, by 
making use of previously-established results obtained in \cite{clpst}, 
which themselves are based on the consistent $S^4$ reduction of
eleven-dimensional supergravity found in \cite{nasvarvan1,nasvarvan2}.  

    From the results in \cite{clpst}, adapted to our conventions and 
the situation where
the $SO(4)$ gauge group considered there is allowed to become 
non-compact, the bosonic reduction ansatz from $D=10$ to $D=7$ becomes
\bea
d\hat s_{10} ^2 &=& \Phi^{3/16}\,\Delta^{1/4}\, 
( ds_7^2 + \ft12 \bar g^{-2}\,\Phi^{-1/2}\, \Delta^{-1} \, 
M^{-1}_{AB}\, D\mu^A\, D\mu^B)\,,\nn\\
\hat F_\3 &=& \ft14 \bar g^{-2}\, \Delta^{-2} 
\, \ep_{A_1\cdots A_4}\, \mu^{B_1}\, \mu^{B_2}\,\eta^{A_1 C_1}\, 
\eta^{A_2 C_2}\,M_{C_1 B_1}\, D M_{C_2 B_2}\wedge D\mu^{A_3}\wedge 
D\mu^{A_4}\nn\\
&& - \ft12 \bar g^{-2}\, U\, \Delta^{-2}\, W +  
\ft12 \bar g^{-1}\, \Delta^{-1}\, \ep_{A_1\cdots A_4}\, 
M_{A_1 B}\, \mu^B \, F_\2^{A_2 A_3}\wedge D\mu^{A_4} + 
 H_\3\,,\nn\\
e^{\hat\phi} &=& \Phi^{5/8}\, \Delta^{-1/2}\,,\label{d10ans}
\eea
where $\mu^A$ are coordinates on $\R^4$, subject to the constraint
\be
\eta_{AB}\, \mu^A\, \mu^B =1\,,\label{surface}
\ee
and 
\bea
D\mu^A &=& d\mu^A + 2 \bar g\, A_{\1}^A{}_B\, \mu^B\,,\nn\\
\Delta &=& M_{A}\, \mu^A\, \mu^B\,,\qquad U = 2 M_{AB}\, M_{CD}\, 
\mu^A\, \mu^C\, \eta^{BD} - \Delta\, M_{AB}\, \eta^{AB}\,,\nn\\
W &=& \ft16 \ep_{A_1 \cdots A_4}\, \mu^{A_1}\, D\mu^{A_2}\wedge 
D\mu^{A_3}\wedge D\mu^{A_4}\,.\label{udw}
\eea
(We are now suppressing the ``bar'' that we placed on the $SO(4)$ gauge
indices $\bar A$ in the previous sections.)  The ansatz (\re{d10ans})
describes the embedding of the truncated $N=2$ gauged
seven-dimensional theory, with gauge group $SO(4)$ or $SO(2,2)$,
depending upon the choice made for $\eta_{AB}$.

  Let us now specialise to the truncation that gave us the Salam-Sezgin 
theory in six dimensions. Using the coordinates introduced in
equation (\ref{coords}), the quantities
$\Delta$, $U$ and $W$ defined in (\ref{udw}) become
\bea
\Delta &=& \cosh2\rho\,,\qquad U=2\,, \nn\\
W &=& -\ft12 \sinh2\rho \, d\rho\wedge (d\a-\bar g\, A)\wedge
(d\beta + \bar g\, A)\,.
\eea
Combining the $D=10$ to $D=7$ reduction with the Kaluza reduction to
$D=6$ given in section \ref{kaluzasec}, we therefore arrive at the 
following reduction ansatz that describes the embedding of the 
Salam-Sezgin theory into ten-dimensional type I supergravity:
\bea
d\hat s_{10}^2 &=& (\cosh2\rho)^{1/4}\, 
\Big[ e^{-\ft14 \phi}\, ds_6^2 + e^{\ft14\phi}\, dz^2 \nn\\
&&\qquad \qquad\qquad  + 
\ft12 \bar g^{-2}\,e^{\ft14\phi}\, ( d\rho^2 + \fft{c^2}{\cosh2\rho}\, 
(d\a -\bar g\, A)^2 + \fft{s^2}{\cosh2\rho}\, 
(d\beta + \bar g\, A)^2 )\Big]\,,\nn\\
\hat F_\3 &=& \fft{s\, c}{\bar g^2\, (\cosh2\rho)^2}\, 
d\rho\wedge (d\a-\bar g\, A_\1)\wedge (d\beta +\bar g\, A_\1)\nn\\
&& +
\fft{1}{2\bar g\, \cosh2\rho}\, F_\2\wedge [c^2\, (d\a-\bar g\, A_\1) -
s^2\, (d\beta + \bar g\, A_\1)] +  H_\3\,,\nn\\
e^{\hat \phi} &=& (\cosh2\rho)^{-1/2}\, e^{-\ft12\phi}\,,\label{ans3}
\eea
where we have defined $c=\cosh\rho$, $s=\sinh\rho$.

    It is straightforward to verify, by direct substitution of the
reduction ansatz (\ref{ans3}) into the equations of motion following
from the Lagrangian
\be
{\cal L}_{10} = \hat R\, {\hat *\oneone} - \ft12 {\hat *d\hat\phi}\wedge
d\hat\phi - \ft12 e^{-\hat\phi}\, {\hat * \hat F_\3}\wedge \hat F_\3
\ee
for the bosonic sector of ten-dimensional type I supergravity, that one
indeed obtains the bosonic equations of motion for the Salam-Sezgin
six-dimensional theory, which follow from (\ref{d6boslag}).

   The metric reduction ansatz in (\ref{ans3}) takes a somewhat more elegant
form in the string frame, related to the Einstein frame by 
$d\hat s_{\rm str}^2 = e^{\ft12\hat\phi}\, d\hat s_{10}^2$:
\be
d\hat s_{\rm str}^2 = 
e^{-\ft12 \phi}\, ds_6^2 +dz^2 
+ 
\ft12 \bar g^{-2}\, \Big( d\rho^2 + \fft{c^2}{\cosh2\rho}\, 
(d\a -\bar g\, A)^2 + \fft{s^2}{\cosh2\rho}\, 
(d\beta + \bar g\, A)^2 \Big)\,.\label{metans4}
\ee

    The ten dimensional string coupling constant is given by $g_s =
e^{\hat\phi}$, and so it goes to zero at large distances $\rho$ in
the internal directions.  Naively at least, the ratio $G_{10}/G_6$ of
the Newton constants in ten and seven dimensions is given by
\be
\fft{G_{10}}{G_6} = \fft{\pi^2}{\sqrt 2 \bar g^{3}}\, 
\int dz\, \int_0^\infty d\rho \sinh2\rho\,,
\ee
and so the diverging $\rho$ integration leads to a vanishing 
six-dimensional gravitational constant.  This is the customary 
feature that one encounters when the internal space has infinite 
volume \cite{gibhul}.

   Now that we have obtained the explicit formulae describing the
embedding of the Salam-Sezgin supergravity in ten-dimensional
supergravity, we can uplift any solution of the Salam-Sezgin theory.
In fact since it was shown in \cite{gibpop} that there exists a
consistent Pauli reduction of the Salam-Sezgin theory on $S^2$, to
give a four-dimensional chiral $N=1$ supergravity, it follows that by
combining this with our new results we can obtain a consistent embedding
of this four-dimensional theory in the ten-dimensional type I
supergravity.  A solution of particular interest is the Salam-Sezgin
(Minkowski)$_4 \times S^2$ ground state, which is given by
\cite{salsez}
\bea
ds_6^2 &=& dx^\mu\, dx_\mu + \fft1{8 \bar g^2}\, (d\theta^2 + 
\sin^2\theta\, d\varphi^2)\,,\nn\\
A_\1 &=& -\fft1{2\bar g}\, \cos\theta\, d\varphi\,,\qquad 
H_\3=0\,,\qquad \phi=0\,.
\eea
Using the ten-dimensional string frame ansatz (\ref{metans4}), this lifts 
to give the ten-dimensional solution
\bea
d\hat s_{\rm str}^2 &=& dx^\mu\, dx_\mu + \fft1{8 \bar g^2}\, (d\theta^2 + 
\sin^2\theta\, d\varphi^2) + dz^2 \nn\\
&&+ \fft1{2\bar g^2}\, 
\Big( d\rho^2 + \fft{c^2}{\cosh2\rho}\, 
(d\a + \ft12 \cos\theta\, d\varphi)^2 +  \fft{s^2}{\cosh2\rho}\, 
(d\beta - \ft12 \cos\theta\, d\varphi)^2\Big)\,.
\eea
The ten-dimensional 3-form and dilaton are given by (\ref{ans3}).

  In the large-$\rho$ limit, the solution approaches
\bea
d\hat s_{\rm str}^2 &=& dx^\mu\, dx_\mu + dz^2 + \fft{d\rho^2}{2\bar g^2} 
\nn\\
&&+
\fft1{8 \bar g^2}\,\Big( d\theta^2 + 
\sin^2\theta\, d\varphi^2 + (d\a-d\beta + \cos\theta\, d\varphi)^2
 + (d\a+d\beta)^2\Big)\,,\nn\\
\hat F_\3 &=& \fft1{8 \bar g^2}\, \sin\theta\, d\theta\wedge d\varphi
\wedge (d\a-d\beta + \cos\theta\, d\varphi)\,.\label{vacsol}
\eea
This asymptotic limit is a well-known exact solution of string theory,
sometimes called the linear-dilaton vacuum, which arises as the
near-horizon geometry of the NS5-brane.  Specifically, the solution is
defined on ${\Bbb E}^{3,1} \times S^1\times S^1 \times \R\times S^3$,
with coordinates $(x^\mu, z, \a+\beta,\rho, \theta,\varphi,\a-\beta)$
respectively.  The NS-NS 3-form $\hat F_\3$ is proportional to the
volume form of the $S^3$, and the dilaton $\hat\phi$ is a linear
function of the coordinate $\rho$, namely $\hat\phi \longrightarrow
-\rho$.  The coordinates $(x^\mu, z, \a+\beta)$ span the NS5-brane
world-volume (which is therefore wrapped over a $T^2$), and the
coordinates $(\rho,\theta,\varphi,\a-\beta)$ cover the transverse
space. Note, however, that in the case of the NS5-brane, moving towards the
horizon corresponds to approaching the strong-coupling region (\ie 
$\hat\phi$ increasing), whereas
in the asymptotic limit we are considering here, the coupling decreases to
zero.  In the NS5-brane case, decreasing coupling corresponds to
moving away from the horizon. In the exact solution we have obtained here,
$e^{\hat\phi}$ is bounded above by 1.

   Although we arrived at the embedding of the Salam-Sezgin theory 
in ten dimensions via an eleven-dimensional and type IIA supergravity
reduction, we can equally well view it as a reduction of type IIB
supergravity, since this shares the same type I supergravity common
sector.  We can then perform an S-duality transformation, so that
the 3-form $\hat F_\3$ in (\ref{ans3}) becomes the R-R 3-form of the type
IIB theory.  In the process, the sign of the dilaton in (\ref{ans3})
will be reversed. 

  Within this type IIB framework, the asymptotic geometry of the solution
(\ref{vacsol}) becomes the near-horizon geometry of the D5-brane.  In
this limit the coupling now becomes strong rather than weak.

\section{Conclusions}

    In this paper we have shown that the six-dimensional $N=(1,0)$ 
gauged Einstein-Maxwell supergravity of Salam and Sezgin can be 
obtained via a consistent reduction from type I supergravity in ten 
dimensions.  The embedding involves a reduction on the three-dimensional 
non-compact space ${\cal H}^{2,2}$ times $S^1$, where ${\cal H}^{2,2}$ is
the quadric 
\be
\mu_1^2 + \mu_2^2 -\mu_3^2 -\mu_4^2 =1
\ee
in Euclidean space ${\Bbb E}^4$, followed by a reduction on $S^1$.
The metric on ${\cal H}^{2,2}$ is positive definite, and is conformal
to the metric induced from the Euclidean metric on ${\Bbb E}^4$.  The
reduction ansatz for the bosonic fields is given by (\ref{ans3}).
Upon substitution into the equations of motion of ten-dimensional type
I supergravity, one obtains the equations of motion of the
Salam-Sezgin theory.
  
   The reduction procedure involved performing consistent truncations
as well as the consistent reductions.  If one elects not to perform the
truncations, then one obtains larger, and non-chiral, gauged 
supergravities in six dimensions, which contain the Salam-Sezgin theory
upon truncation.  In particular, we exhibited an $SO(2,2)$ gauged $N=(1,1)$
supergravity in six dimensions, that arose from the $S^1$ reduction of
seven-dimensional $N=2$ gauged $SO(2,2)$ supergravity.  We could also
obtain an $N=(2,2)$ gauged supergravity in six dimensions if we 
did not make the truncation from $N=4$ to $N=2$ in seven dimensions.

   Since the Salam-Sezgin theory is embedded in ten-dimensional 
type I supergravity, it follows that our construction can equally well
be viewed as a reduction of the heterotic theory, or, via an S-duality
transformation, as a reduction of type IIB supergravity in which the
R-R rather than the NS-NS 3-form is non-vanishing.  We can also, of
course, trivially lift the embedding (viewed as type I within type IIA) to
an embedding within eleven-dimensional supergravity.

   Having obtained the Salam-Sezgin theory via a consistent dimensional
reduction, it follows that any of its solutions can be lifted to give 
a solution in $D=10$ or $D=11$.  It is striking that the lift of the
(Minkowski)$_4\times S^2$ ground state to type I supergravity is 
asymptotic to the exact linear-dilaton solution of string theory, \ie
the near-horizon geometry of the NS5-brane.  By S-duality, it can be
viewed instead as the near-horizon geometry of the D5-brane.

\section*{Acknowledgments}

    We are grateful to Klaus Behrndt, Rahmi G\"uven, Jim Liu, Hong
L\"u, Krzysztof Pilch, Toine van Proyen, Fernando Quevedo, Ergin
Sezgin and Paul Townsend for discussions. We thank the Benasque Center
for Science, and M.C. and C.N.P. thank the Cambridge Relativity and
Gravitation Group and the organisers of the {\sl Cosmological
Perturbations on the Brane} workshop, for hospitality during the
course of this work.


\end{document}